\documentclass[useAMS,usenatbib]{mn2e}

\usepackage{graphicx}
\usepackage{natbib}
\usepackage{amsmath}

\def\lesssim{\mathrel{\hbox{\rlap{\hbox{\lower5pt\hbox{$\sim$}}}\hbox{$<$}}}}
\def\gtrsim{\mathrel{\hbox{\rlap{\hbox{\lower5pt\hbox{$\sim$}}}\hbox{$>$}}}}

\def\CKL{C_{\textrm{KL}}}
\def\CKLmax{C_{\textrm{KL}, \max}}
\def\CKLmin{C_{\textrm{KL}, \min}}
\def\crit{\textrm{crit}}
\def\ekm{\textrm{EKM}}
\def\epsoct{\epsilon_{\textrm{oct}}}
\def\Hhat{\hat{\mathcal{H}}}
\def\Hhatquad{\hat{\mathcal{H}}_{\textrm{q}}}

\def\Hquad{\mathcal{H}_{\textrm{q}}}

\def\in{\textrm{in}}

\def\KL{\textrm{KL}}
\def\Msun{M_{\odot}}

\def\out{\textrm{out}}
\def\phiqcrit{\phi_{q, \textrm{crit}}}
\def\Pin{P_{\textrm{in}}}
\def\Pout{P_{\textrm{out}}}
\def\tauekm{\tau_{\textrm{EKM}}}

\def\tekm{t_{\textrm{EKM}}}
\def\tkl{t_{\textrm{KL}}}
\def\tklflip{t_{\textrm{KL}, i = 90^{\circ}}}

\title[Timescales of Kozai-Lidov oscillations]{Timescales of Kozai-Lidov
oscillations at quadrupole and octupole order in the test particle limit}
\author[Antognini]{J.~M.~O.~Antognini$^{1, 2}$
\\
$^1$ Department of Astronomy, The Ohio State University, Columbus, Ohio
43210, USA\\
$^2$ Center for Cosmology and Astro-Particle Physics, The Ohio State
University, Columbus, Ohio 43210, USA\\
E-mail: antognini@astronomy.ohio-state.edu}

\begin{document}

\maketitle

\begin{abstract}

Kozai-Lidov (KL) oscillations in hierarchical triple systems have found
application to many astrophysical contexts, including planet formation, type
Ia supernovae, and supermassive black hole dynamics.  The period of these
oscillations is known at the order-of-magnitude level, but dependencies on
the initial mutual inclination or inner eccentricity are not typically
included.  In this work I calculate the period of KL oscillations ($\tkl$)
exactly in the test particle limit at quadrupole order (TPQ).  I explore the
parameter space of all hierarchical triples at TPQ and show that except for
triples on the boundary between libration and rotation, the period of KL
oscillations does not vary by more than a factor of a few.  The exact period
may be approximated to better than 2 per cent for triples with mutual
inclinations between 60$^{\circ}$ and 120$^{\circ}$ and initial
eccentricities less than $\sim$0.3.  In addition, I derive an analytic
expression for the period of octupole-order oscillations due to the
`eccentric KL mechanism' (EKM).  I show that the timescale for EKM
oscillations is proportional to $\epsoct^{-1/2}$, where $\epsoct$ measures
the strength of octupole perturbations relative to quadrupole perturbations.

\end{abstract}

\section{Introduction}
\label{sec:intro}

Triple systems are common in the Galaxy, comprising $\sim$10\% of systems in
which the primary is $\sim$1 $\Msun$ \citep{duquennoy91, raghavan10,
tokovinin14, riddle15}.  All observed triples are `hierarchical,' in that
the relative distance between two components of the triple is much smaller
than the relative distance between them and the third.  Such systems are
stable if they are sufficiently hierarchical \citep{mardling99,
mardling01}.\footnote{While stable non-hierarchical triple systems are
possible \citep[e.g.,][]{chenciner00, suvakov13}, they may require fine
tuning to form and have never been observed in nature.}  

In general, if the tertiary is highly inclined relative to the inner binary,
the eccentricity of the inner binary will undergo oscillations, known as
Kozai-Lidov (KL) oscillations \citep{lidov62, kozai62}.  KL oscillations
have been invoked in many contexts to explain a wide variety of phenomena
such as the formation of hot Jupiters \citep{wu03, wu07, fabrycky07, naoz11,
naoz12, petrovich15a}, the formation of blue stragglers \citep{perets09,
naoz14a}, the merger of WD-WD binaries \citep{thompson11, katz11}, the
merger of supermassive and intermediate-mass black holes \citep{miller02b,
blaes02, wen03}, the distribution of dark matter around supermassive black
hole binaries \citep{naoz14b}, and as a source of unique gravitational wave
signals \citep{miller02a, gould11, seto13, antonini14, antognini14, bode14}. 

KL oscillations are a secular phenomenon, occurring on timescales much
longer than the orbital periods.  It is therefore possible to average the
motions of the individual stars over their orbits and study only the secular
changes to the orbital elements.  If there is a large mass ratio in the
inner binary then on even longer timescales the strength of the KL
oscillations (i.e., the maximum eccentricity reached) will vary
\citep{ford00, katz11, lithwick11, naoz13a}.  These variations have been
termed the `eccentric KL mechanism' (EKM), and in some cases can cause the
inner binary to pass through an inclination of 90$^{\circ}$ with respect to
the outer binary in a `flip' from prograde to retrograde or vice versa.
During a flip the eccentricity of the inner binary can be driven to
extremely large values because the strength of KL oscillations is very
sensitive to the mutual inclination, with arbitrarily strong oscillations
occurring as the inclination approaches 90$^{\circ}$ exactly in the test
particle limit.  Although EKM oscillations do not occur when the two stars
of the inner binary are of equal mass, mass loss from one of the stars in
the course of stellar evolution can induce EKM oscillations
\citep{shappee13b, michaely14}.  EKM oscillations and flips have generally
been studied in the context of hierarchical triples, but flips occur over a
wider range of parameter space in both 2+2 quadruples \citep{pejcha13} and
3+1 quadruples \citep{hamers15}.

Because the extreme eccentricity oscillations that occur during a flip can
affect the evolution of the objects in the inner binary, the timescale for
EKM oscillations is another important quantity.  Yet no derivation of the
timescale for EKM oscillations has appeared in the literature, although
several studies have asserted that $\tekm \sim \tkl / \epsoct$ is a
plausible timescale \citep[e.g.,][]{katz11, naoz13b, li15b}, where
$\epsoct$ measures the strength of the octupole order term relative to the
quadrupole order term of the Hamiltonian (see equation~\ref{eq:epsoct_def}
for a definition).  I show that $\tekm \sim \tkl / \sqrt{\epsoct}$. 

In Section~\ref{sec:basicequations}, I present the basic parameters and
equations that govern a hierarchical three-body system.  In
Section~\ref{sec:derivation}, I then derive the period of KL oscillations.
In Section~\ref{sec:survey}, I explore how the period varies over the
parameter space and in Section~\ref{sec:approx}, I provide an approximation
to the exact period.  In Section~\ref{sec:ekm}, I treat the period
of EKM oscillations and derive the corrected timescale.  I conclude in
Section~\ref{sec:conclusions}.

To perform the calculations in this paper I wrote the \texttt{kozai} Python
module.  This module can evolve hierarchical triple systems in the secular
approximation up to hexadecapole order using either the Delaunay orbital
elements or the eccentricity and angular momentum vectors.  I have released
this code under the MIT license and it is available at
https://github.com/joe-antognini/kozai. 

\section{Basic equations}
\label{sec:basicequations}

\subsection{Notation}
\label{subsec:notation}

Throughout this paper orbital properties referring to the inner and outer
binary are labelled with a subscript 1 and 2, respectively.  The masses of
the two components of the inner binary are $m_1$ and $m_2$, and the mass of
the tertiary is $m_3$.

We will often refer to the orbital parameters using Delaunay's elements: the
mean anomalies, $l_x$; the arguments of periapsis, $g_x$, and the longitudes
of ascending nodes, $h_x$, where $x = 1$ or 2 and refers to the inner or
outer binary, respectively.  Their conjugate momenta are
\begin{equation}
\label{eq:L1}
L_1 = \frac{m_1 m_2}{m_1 + m_2} \sqrt{G (m_1 + m_2) a_1},
\end{equation}
\begin{equation}
\label{eq:L2}
L_2 = \frac{m_3 (m_1 + m_2)}{m_1 + m_2 + m_3} \sqrt{G (m_1 + m_2 + m_3) a_2},
\end{equation}
\begin{equation}
G_x = L_x \sqrt{1 - e_x^2},
\end{equation}
and
\begin{equation}
H_x = G_x \cos i_x.
\end{equation}
Delaunay's elements form a set of canonical variables.  Note that $G_1$ and
$G_2$ are the angular momenta of the inner and outer binaries, respectively.
We furthermore define the reduced angular momentum
\begin{equation}
j_x^2 \equiv 1 - e_x^2.
\end{equation}
The angular momentum of an orbit may thus be written $G_x = L_x j_x$. 

\subsection{The Hamiltonian}
\label{subsec:hamiltonian}

If a three-body system is sufficiently hierarchical, its Hamiltonian may be
considered to be that of two isolated binaries (the inner binary, consisting
of the two closest bodies, and the outer binary, consisting of the distant
body plus the inner binary taken as a point mass) plus a perturbative
interaction term:
\begin{equation}
\mathcal{H} = \frac{G m_1 m_2}{2 a_1} + \frac{G(m_1 + m_2) m_3}{2a_2} +
\mathcal{H}_{\textrm{pert}}.
\end{equation}
This interaction term captures the change in the orbital motion of each
binary in the tidal field of the other.  Because we are assuming that the
triple is hierarchical, the semi-major axis ratio, $\alpha = a_1/a_2$, is a
small parameter that we can use to expand the perturbative component of the
Hamiltonian in a multipole expansion \citep{harrington68},
\begin{equation}
\label{eq:multipole}
\mathcal{H}_{\textrm{pert}} = \frac{G}{a_2} \sum_{j=2}^{\infty} \alpha^j
\mathcal{M}_j \left( \frac{r_1}{a_1} \right)^j \left( \frac{a_2}{r_2}
\right)^{j+1} P_j (\cos \Phi),
\end{equation}
where $P_j$ is the $j^{\textrm{th}}$ Legendre polynomial, $r_x$ is the
distance between the two components of the $x^{\textrm{th}}$ binary, $\Phi$
is the angle between $r_2$ and $r_1$ and $\mathcal{M}_j$ is a mass parameter
defined by
\begin{equation}
\label{eq:mass_parameter}
\mathcal{M}_j = m_1 m_2 m_3 \frac{m_1^{j-1} - (-m_2)^{j-1}}{(m_1 + m_2)^j}. 
\end{equation}

If we are only interested in changes to the orbital elements that occur on
timescales much longer than the orbital periods (so-called `secular'
changes), we must average the Hamiltonian over both mean anomalies.  To do
this while maintaining the canonical structure of the Hamiltonian requires a
technique known as von Zeipel averaging.  The general case for three massive
bodies is quite complicated even at quadrupole order as one must be careful
to include the longitudes of ascending nodes \citep{naoz13a}.  However, the
Hamiltonian simplifies considerably if one component of the inner binary is
taken to be a test particle as this allows one to fix the longitudes of
ascending nodes and eliminate them from the Hamiltonian.  The resulting
double-averaged Hamiltonian at quadrupole order in the test particle limit
is
\begin{equation}
\label{eq:quadrupole}
\mathcal{H}_{\textrm{q}} = C_2 \left[\left(2 + 3e_1^2\right)\left(1 -
3\cos^2 i \right) - 15 e_1^2 \left(1 - \cos^2 i \right) \cos 2 g_1\right],
\end{equation}
where $C_2$ is a constant parameterizing the strength of the quadrupole term
given by
\begin{equation}
\label{eq:c2}
C_2 = \frac{G m_1 m_2 m_3}{16 (m_1 + m_2) a_2 (1 - e_2^2)^{3/2}} \alpha^2.
\end{equation}
The semi-major axes and $e_2$ do not change at quadrupole order, so $C_2$ is
also constant.  We will henceforth refer to the dimensionless Hamiltonian,
\begin{equation}
\label{eq:hhat}
\Hhatquad \equiv \frac{\mathcal{H}_q}{C_2}.
\end{equation}

\subsection{Integrals of motion}
\label{subsec:integrals}

There are no dissipative forces in the problem, so the total energy,
$\Hhat$, remains constant.  Moreover, because no energy is transferred
between the two binaries at quadrupole order, each term of the Hamiltonian
is conserved separately, so $\Hhatquad$ remains constant as well.  

The total angular momentum is also conserved and may be expressed in the
form of the geometrical relation,
\begin{equation}
\label{eq:geometricalrelation}
\cos i = \frac{G_{\textrm{tot}}^2 - G_1^2 - G_2^2}{2 G_1 G_2}.
\end{equation}
This relation is valid in the general case of three massive bodies.  In the
test particle limit the geometrical relation may be approximated by
\begin{equation}
G_{\textrm{tot}} \simeq G_2 + G_1 \cos i.
\end{equation}
Since $G_{\textrm{tot}}$ and $G_2$ are constant, we must have that $G_1 \cos
i$ is constant as well.  Furthermore, $G_1 = L_1 j_1$, and $L_1$ is also
constant, so this requires that $j_1 \cos i$ be constant as well.  We notate
this constant of motion as
\begin{equation}
\label{eq:Theta}
\Theta \equiv (j_1 \cos i)^2.
\end{equation}
In this form, the constant of motion is known as `Kozai's integral' and is
equal to the square of the $z$-component of the reduced angular momentum,
$j_z$ \citep{holman97}.  Kozai's integral implies that the component of
angular momentum perpendicular to the plane of the outer binary is constant.
However, the test particle assumption is crucial to its derivation.  In the
general case of three massive bodies this component of angular momentum is
not conserved, although it is possible to derive a generalized version which
is conserved \citep[e.g.,][]{wen03}.  The generalized Kozai integral is due
to the fact that, as \citet{lidov76} showed, $\Hhatquad$ is independent of
$g_2$, thereby implying that $G_2$ (and hence also $e_2$) must be constant.

Because $\Hhatquad$ only depends on $e_1$, $\cos i$, and $g_1$, and there
are two integrals of motion, $\Hhatquad$ and $\Theta$, there is only one
degree of freedom and so the system is integrable.  Moreover, because these
variables are all bounded, the motion is periodic (with the exception of a
locus of stationary points of measure zero).  The Hamiltonian to quadrupole
order may be expressed as 
\begin{equation}
\label{eq:H}
\Hhatquad = \frac{1}{j_1^2} \left[ (5 - 3j_1^2) (j_1^2 - 3 \Theta) - 15 (1 -
j_1^2)(j_1^2 - \Theta) \cos 2 g_1 \right]
\end{equation}
in terms of $j_1$ and $\Theta$. 

\subsection{Equations of motion}
\label{subsec:eom}

We are interested in the time evolution of the variables $j_1$, $\cos i$,
and $g_1$.  Of these, only $g_1$ is a canonical variable so its time
evolution follows directly from Hamilton's equations:
\begin{equation}
\frac{dg_1}{dt} = \frac{\partial \Hquad}{\partial G_1} = \frac{C_2}{L_1}
\frac{\partial \Hhatquad}{\partial j_1}
\end{equation}
Carrying out the differentiation of equation~(\ref{eq:H}) we find
\begin{equation}
\label{eq:dgdt}
\frac{dg_1}{dt} = \frac{6 C_2}{L_1} \frac{1}{j_1^3} \left[5 \left(\Theta -
j_1^4\right) \left(1 - \cos 2 g_1\right) + 4 j_1^4 \right]
\end{equation}
The variable $j_1$ is related to a canonical variable, $G_1$, by a constant,
so we find its time evolution to be
\begin{equation}
\frac{dj_1}{dt} = \frac{1}{L_1} \frac{\partial \Hquad}{\partial g_1} =
\frac{C_2}{L_1} \frac{\partial \Hhatquad}{\partial g_1}.
\end{equation}
Again carrying out the differentiation of equation~(\ref{eq:H}) we find
\begin{equation}
\label{eq:dedt}
\frac{dj_1}{dt} = \frac{30 C_2}{L_1} \frac{1}{j_1^2} \left(1 - j_1^2\right)
\left(j_1^2 - \Theta \right) \sin 2 g_1.
\end{equation}

The time evolution of the inclination is complicated by the elimination of
nodes from the Hamiltonian.  Due to this procedure the time evolution of the
inclination cannot be recovered from Hamilton's equations directly.
Instead, the inclination must be derived by calculating $j_1$ and solving
the geometrical relation given in equation~(\ref{eq:geometricalrelation}).

\subsection{Libration vs.~rotation}
\label{subsec:libration}

During a KL oscillation, the argument of periapsis of the inner binary may
either rotate or librate.  This is to say, $g_1$ may sweep through the full
range of angles from 0 to $2\pi$ (rotation) or it may be restricted to just
a subset of them (libration).  In the case of libration, the set of
librating trajectories must librate about a fixed point of $g_1$ and $j_1$.
Inspection of equation~(\ref{eq:dedt}) reveals that $j_1$ is stationary only
when $g_1$ takes half- or whole-integer multiples of $\pi$ (recall that
$\Theta < j_1^2$).  Now, inspection of equation~(\ref{eq:dgdt}) reveals that
$g_1$ cannot be stationary at integer multiples of $\pi$.  This implies that
trajectories can only librate about half-integer multiples of $\pi$, so
$g_{1, \textrm{fix}} = \pm \pi/2$ and $j_{1, \textrm{fix}}^2 =
\sqrt{5\Theta/3}$. 

To determine whether a particular system (i.e., a given $\Hhatquad$ and
$\Theta$) librates or rotates we must see whether there exists a physical
solution of equation~(\ref{eq:H}) for $j_1$ when $g_1 = 0$.  Setting $g_1 =
0$ in equation~(\ref{eq:H}) and solving for $j_1^2$ we find
\begin{equation}
j_1^2 = \frac{1}{12}(10 + \Hhatquad + 6 \Theta).
\end{equation}
The critical system on the boundary between libration and rotation will have
a solution for $j_1$ exactly equal to unity and libration will occur if the
only solution for $j_1$ exceeds unity.  Defining the libration constant as
\begin{equation}
\label{eq:librationconst}
\CKL \equiv \frac{1}{12} \left(2 - \Hhatquad - 6 \Theta \right),
\end{equation}
we will have libration if $\CKL < 0$ and rotation if $\CKL > 0$.  This
constant was first presented in \citet{lidov62} and may be calculated
equivalently by
\begin{equation}
\label{eq:katzCK}
\CKL = e^2 \left( 1 - \frac{5}{2} \sin^2 i \sin^2 g_1 \right).
\end{equation}
Note that the condition for rotation then becomes
\begin{equation}
\sin g_1 \leq \sqrt{\frac{2}{5}} \frac{1}{\sin i}.
\end{equation}
Because $\CKL$ naturally parameterizes a dynamical property of the triple,
it is often convenient to work with it instead of $\Hhatquad$ where
possible.  

\section{Derivation of the period of KL oscillations}
\label{sec:derivation}

Because the Hamiltonian at quadrupole order is integrable, the period of KL
oscillations, $\tkl$, may be determined exactly.  The period may be written
\begin{equation}
\label{eq:base_tkl}
\tkl = \oint \, dt = \oint \frac{dt}{dj_1} \, dj_1.
\end{equation}
Solving equation~(\ref{eq:H}) for $\cos 2 g_1$ and rewriting in terms of
$\sin 2 g_1$, we have
\begin{equation}
\label{eq:sin2g1}
\sin 2 g_1 = \left[1 - \left(\frac{3j_1^4 + j_1^2 \left(\Hhatquad - 9 \Theta
- 5\right) + 15 \Theta}{15 \left(1 - j_1^2\right) \left(j_1^2 -
\Theta\right)}\right)^2\right]^{\frac{1}{2}}.
\end{equation}
Substituting equation~(\ref{eq:sin2g1}) into equation~(\ref{eq:base_tkl})
and substituting the result into equation~(\ref{eq:base_tkl}) we find
\begin{multline}
\label{eq:loopintegral}
\tkl = \frac{L_1}{30 C_2} \oint \frac{j_1^2}{(1 - j_1^2)(j_1^2 - \Theta)}\\
\times \left[1 - \left(\frac{3 j_1^4 + j_1^2(\Hhatquad - 9 \Theta - 5) + 15
\Theta}{15(1 - j_1^2)(j_1^2 - \Theta)} \right)^2 \right]^{-\frac{1}{2}} \,
dj_1.
\end{multline}
We note that this integral may be rewritten in terms of incomplete elliptic
integrals of the first kind, but we do not do so here because it complicates
the expression considerably.

The integral in equation~(\ref{eq:loopintegral}) proceeds from the maximum
value of $j_1$ to the minimum value of $j_1$ and back again to the maximum
value of $j_1$, so we may instead integrate from $j_{\min}$ to $j_{\max}$
and multiply by two.  Eliminating $\Hhatquad$ in favor of $\CKL$ by making
use of equation~(\ref{eq:librationconst}), and rearranging, we have
\begin{multline}
\label{eq:kl_period}
\tkl = \frac{L_1}{15 C_2} \int_{j_{\min}}^{j_{\max}} \frac{1}{(1 - j_1^2)}\\
\times
\left[\left(1 - \frac{\Theta}{j_1^2}\right)^2 - \left(\frac{1}{5} -
\frac{\Theta}{j_1^2} + \frac{4}{5} \frac{\CKL}{1 - j_1^2} \right)^2
\right]^{-\frac{1}{2}} \, d j_1.
\end{multline}
Eccentricity maxima ($j_{\min}$) occur for $g_1 = \pm \pi/2$.  Eccentricity
minima ($j_{\max}$) also occur at $g_1 = \pm \pi/2$ in the case of libration
but occur at $g_1 = 0$ or $\pi$ in the case of rotation.  We may therefore
solve for $j_{\min}$ and $j_{\max}$ by substituting the appropriate values
of $g_1$ into equation~(\ref{eq:H}) and solving for $j_1$.  We therefore
have
\begin{equation}
\label{eq:epsmin}
j_{\min} = \sqrt{\frac{1}{6} \left(\zeta - \sqrt{\zeta^2 - 60 \Theta}
\right)}
\end{equation}
\begin{equation}
\label{eq:epsmaxlib}
j_{\max} =  \sqrt{\frac{1}{6} \left(\zeta + \sqrt{\zeta^2 - 60
\Theta}\right)}, \quad \CKL < 0
\end{equation}
\begin{equation}
\label{eq:epsmaxrot}
j_{\max} = \sqrt{1 - \CKL}, \quad \CKL > 0.
\end{equation}
where we have defined
\begin{equation}
\zeta \equiv 3 + 5 \Theta + 2 \CKL.
\end{equation}
For convenience, we define the integral in equation~(\ref{eq:kl_period}) to
be $f(\CKL, \Theta)$ such that
\begin{equation}
f(\CKL, \Theta) \equiv \frac{15 \tkl C_2}{L_1}.
\end{equation}
Having calculated the limits of integration, we can now use
equation~(\ref{eq:kl_period}) to calculate the period of KL oscillations to
quadrupole order in the test particle limit for any hierarchical triple. 

\section{A brief survey of parameter space}
\label{sec:survey}

We now turn to a brief exploration of the range of values that the integral
in equation~(\ref{eq:kl_period}) may take.  The overall timescale for KL
oscillations is determined by the coefficient before the integral, which we
present in more detail in Section~\ref{subsec:kltimescale}.  The integral,
however, depends on only two parameters describing the triple: $\Hhatquad$
and $\Theta$, or equivalently, $\CKL$ and $\Theta$.  Thus, once the
timescale of KL oscillations has been set, only two degrees of freedom
remain.

What values may $\Hhatquad$, $\CKL$, and $\Theta$ take?  It is easy to see
from equation~(\ref{eq:Theta}) that $0 \leq \Theta \leq 1$ since both $j_1$
and $\cos i$ are bounded by 0 and 1.  Moreover, it is clear from
equation~(\ref{eq:katzCK}) that
\begin{equation}
-\frac{3}{2} \leq \CKL \leq 1
\end{equation}
since all the terms are bounded by 0 and 1.  From the bounds on $\Theta$ and
$\CKL$, we can conclude from equation~(\ref{eq:librationconst}) that the
bounds on $\Hhatquad$ are
\begin{equation}
\label{eq:Hrange}
-10 \leq \Hhatquad \leq 20.
\end{equation}
However, the limits on $\Hhatquad$ and $\Theta$ are not independent.  In the
case of $g_1 = 0$, the requirement that $\Theta \leq j_1^2$ implies that
\begin{equation}
\label{eq:ThHrange}
-10 + 6 \Theta \leq \Hhatquad \leq 20.
\end{equation}
This, in turn, translates to the requirement in $\CKL$ that
\begin{equation}
\label{eq:ThCKLrange}
\CKL \leq 1 - \Theta.
\end{equation}

In order for equation~(\ref{eq:kl_period}) to have a solution, the square
roots in equations~(\ref{eq:epsmin}),~(\ref{eq:epsmaxlib}),
and~(\ref{eq:epsmaxrot}) must exist.  The existence of the inner square root
in equation~(\ref{eq:epsmin}) requires that
\begin{equation}
\CKL \geq -\frac{1}{2} \left(5 \Theta - 2 \sqrt{15 \Theta} + 3\right).
\end{equation}
This requirement is always satisfied in the case of rotation ($\CKL > 0$).
In the case of libration ($\CKL < 0$), this requirement may instead be
written in terms of $\CKL$ as
\begin{equation}
\label{eq:full_criterion}
\Theta \leq \frac{1}{5} \left(3 - 2 \sqrt{-6 \CKL} - 2 \CKL \right), \quad
\CKL \leq 0. 
\end{equation}

In the case of libration, the square root in equation~(\ref{eq:epsmaxlib})
exists everywhere that the square root in equation~(\ref{eq:epsmin}) does,
so the existence of this square root adds no new constraints.  In the case
of rotation, the requirement that the square root in
equation~(\ref{eq:epsmaxrot}) exist is satisfied by the same condition set
in equation~(\ref{eq:ThCKLrange}).

Equation~(\ref{eq:full_criterion}) implies that there is a critical inclination,
below which librating KL oscillations do not occur.  Taking $\CKL = 0$ we
recover the usual critical inclination of $\cos i \geq \sqrt{3/5}$. 
Although we do not provide an explicit derivation, we note that the
criterion in equation~(\ref{eq:full_criterion}) can also be arrived at by
requiring that 
\begin{equation}
\frac{d^2 j_1}{dt^2} < 0
\end{equation}
when $j_1$ is at a maximum.  In other words, KL oscillations occur when the
minimum eccentricity is an unstable equilibrium.

\begin{figure}
\centering
\includegraphics[width=8cm]{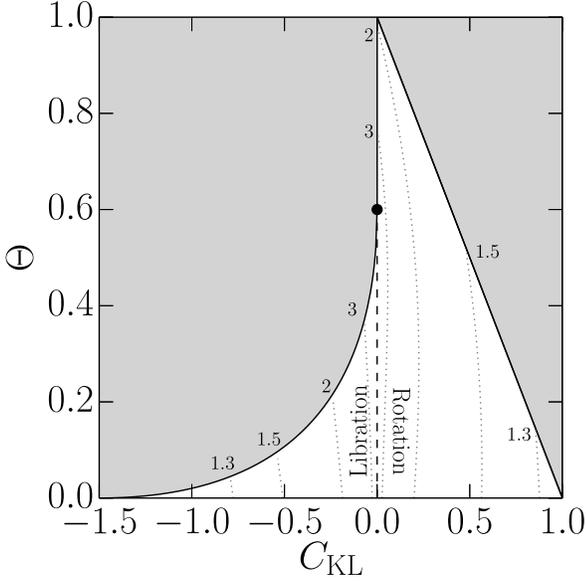}

\caption{Variation in the period of KL oscillations over all of parameter
space.  The contours (dotted lines) show different values of $f(\CKL,
\Theta)$ (i.e., the integral in equation~\ref{eq:kl_period}).  The period
varies only by a factor of a few except very near the boundary between
rotation and libration (dashed line), where it diverges.  The gray regions
indicate where KL oscillations are not possible.  The large dot at $\CKL =
0$, $\Theta = 3/5$ marks the largest value of $\Theta$ for which libration
is possible.}

\label{fig:pspace}
\end{figure}

Knowing now the region of parameter space in which KL oscillations occur, we
can numerically integrate the integral in equation~(\ref{eq:kl_period}) over
the entire parameter space.  The results of this procedure are presented in
Figure~\ref{fig:pspace}.  Except for a narrow strip of parameter space
centered around the boundary between rotation and libration ($\CKL = 0$) the
integral only varies by a factor of a few.  Near the rotation-libration
boundary the integral diverges and KL oscillations have arbitrarily large
periods.  Figure~\ref{fig:pspace} also indicates that the period of KL
oscillations depends most strongly on $\CKL$ and only weakly on $\Theta$.

\section{Approximations}
\label{sec:approx}

\subsection{The timescale of KL oscillations}
\label{subsec:kltimescale}

So long as the integral in equation~(\ref{eq:kl_period}) is of order unity,
the period of KL oscillations will be given by the coefficient before the
integral to within an order of magnitude:
\begin{equation}
\label{eq:kl_timescale}
\tkl \simeq \frac{L_1}{15 C_2}.
\end{equation}
Substituting equations~(\ref{eq:L1}) and~(\ref{eq:c2}) and noting that we
are working in the test particle limit so $m_2 \to 0$, we have the timescale
in terms of the semi-major axes, masses, and eccentricities:
\begin{equation}
\tkl \simeq \frac{16}{15} \left( \frac{a_2^3}{a_1^{3/2}} \right)
\sqrt{\frac{m_1}{G m_3^2}} \left(1 - e_2^2 \right)^{3/2}.
\end{equation}
This timescale may be expressed more elegantly in terms of the periods of
the inner and outer orbits, $\Pin$ and $\Pout$, respectively, by making use
of Kepler's law:
\begin{equation}
\tkl \simeq \frac{8}{15 \pi} \left(1 + \frac{m_1}{m_3} \right) \left(
\frac{\Pout^2}{\Pin} \right) \left(1 - e_2^2\right)^{3/2}.
\end{equation}
This is the form of the KL period that typically appears in the literature,
but with an additional mass term and numerical coefficient.  The mass term
implies that KL oscillations lengthen indefinitely as the tertiary
approaches zero mass.  In the case of a massive tertiary but a test particle
primary and secondary (e.g., a WD-WD binary orbiting a SMBH), the period of
KL oscillations approaches a constant value.  Note that in the case of an
equal mass primary and tertiary, neglecting the numerical coefficient will
lead to an overestimate of the period of KL oscillations by a factor of
nearly three.

\subsection{High inclination, low eccentricity triples}
\label{subsec:highinc}

In most cases of interest in astronomy, the inner binary of a hierarchical
triple starts with a low to moderate eccentricity.  Moreover, KL
oscillations are strongest (and therefore most interesting) when the
tertiary is at high inclination.  It is therefore worth finding an
approximation to $\tkl$ in the high inclination, low initial eccentricity
limit.  In this limit, we have $\Theta \to 0$ and $\CKL \to 0$ and
equation~(\ref{eq:kl_period}) may be solved exactly:
\begin{equation}
f(\CKL, \Theta) \simeq \left. \frac{5}{4 \sqrt{6}} \ln \left( \frac{1 +
j_1}{1 - j_1} \right) \right|^{j_{\max}}_{j_{\min}}
\end{equation}
We also have in this limit that $j_{\min} \ll 1$ and $1 - j_{\max} \ll 1$ so
that
\begin{equation}
f(\CKL, \Theta) \simeq \frac{5}{4 \sqrt{6}} \ln \left( \frac{2}{1 -
j_{\max}} \right).
\end{equation}
where
\begin{equation}
j_{\max} \simeq 1 + \frac{\CKL}{3}
\end{equation}
for libration ($\CKL < 0$), and 
\begin{equation}
j_{\max} \simeq 1 - \frac{\CKL}{2}.
\end{equation}
for rotation ($\CKL > 0$).  

The dependence of $f(\CKL, \Theta)$ on $\Theta$ is non-trivial to
approximate from first principles.  After experimenting with several
functional forms, we found that $f(\CKL, \Theta)$ varies most closely with
$(1 - \Theta)$.  If there is a $\Theta$ dependence both inside and outside
the logarithm, we then expect $f(\CKL, \Theta)$ to take the form
\begin{equation}
f(\CKL, \Theta) \approx \frac{5}{4 \sqrt{6}} \ln \left(\frac{a (1 -
\Theta)^b}{\CKL} \right) \left(1 - \Theta \right)^c,
\end{equation}
where $a$, $b$, and $c$ are fitting parameters.  We fit numerical
integrations of $f(\CKL, \Theta)$ to this form over the range $0 \leq \Theta
\leq 0.25$, $-0.1 \leq \CKL \leq 0.1$ and find the remarkably good fit,
\begin{multline}
\label{eq:approx}
\tkl \approx \frac{1}{3\pi} \sqrt{\frac{2}{3}} \left(1 + \frac{m_1}{m_3}
\right) \left(\frac{P_{\textrm{out}}^2}{P_{\textrm{in}}} \right) \left(1 -
e_2^2\right)^{3/2} \\
\times \ln \left( \frac{9.42 (1 - \Theta)^{2.36}}{\CKL} \right) \left( 1 -
\Theta \right)^{-1.53}.
\end{multline}
We attempted to add several auxiliary parameters but found that they did not
substantially improve the fit. 

\begin{figure}
\centering
\includegraphics[width=8cm]{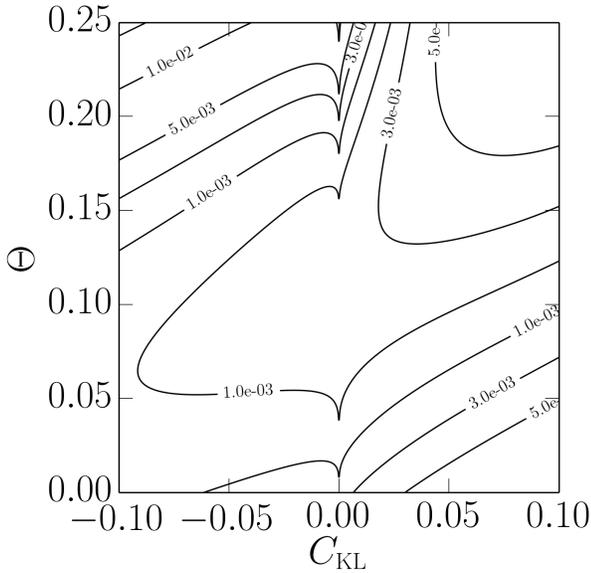}

\caption{Residuals for the approximation in equation (\ref{eq:approx}) to
the period of KL oscillations in the high inclination, low eccentricity
limit.  The approximation is correct to within 2\% at all points in this
range and typically does much better.  This range of $\CKL$ and $\Theta$
corresponds to triples in which the inner binary has an eccentricity $e_1
\lesssim 0.3$ and the inclination is $i \gtrsim 60^{\circ}$.}

\label{fig:approx}
\end{figure}

The approximation provided in equation~(\ref{eq:approx}) fits the true value
of $f(\CKL, \Theta)$ to within 2\% over the range sampled, and over the vast
majority of the range sampled the residuals are less than 0.3\%.  This is
therefore an appropriate formula to use for triples in which the inner
binary has an eccentricity $e_1 \lesssim 0.3$ and an inclination $60^{\circ}
\lesssim i \lesssim 120^{\circ}$.  (Note that the triple need only have high
inclination and low eccentricity at some point in the KL cycle for the
approximation to be valid.)  Contours of the residuals of
equation~(\ref{eq:approx}) are shown in Figure~\ref{fig:approx}.

\section{The eccentric KL mechanism}
\label{sec:ekm}

If the two masses of the inner binary are not equal and the outer orbit has
non-zero eccentricity, the next term in the expansion of the Hamiltonian,
the octupole order term, becomes dynamically significant.  This term leads
to changes to the orbital parameters of the outer orbit that are slow
relative to individual KL oscillations.  These long-term changes can cause
the inner orbit to eventually pass through an inclination of 90$^{\circ}$.
During these orbital flips, the nearly perpendicular inclination leads to strong KL
oscillations which drive the inner binary to extremely large eccentricities.
For this reason, the dynamical effect of the octupole term has been called
the `eccentric KL mechanism' (EKM) \citep[e.g.,][]{lithwick11}.  

The introduction of the octupole term breaks the integrability of the
Hamiltonian.  Consequently, neither $\CKL$ or $\Theta$ remain constants of
the motion.  Furthermore, in the test particle limit at quadrupole order it
is possible to eliminate the longitude of ascending node, $\Omega$, from the
Hamiltonian.  At octupole order either this parameter or $g_2$ necessarily
enters into the equations of motion.  In this section we will follow the
analysis of \citet{katz11} and work in terms of the longitude of ascending
node of the eccentricity vector, $\Omega_e$, defined such that $\textbf{e} =
e (\sin i_e \cos \Omega_e, \sin i_e \sin \Omega_e, \cos i_e)$, and
$\textbf{e}$ points toward periapsis of the inner binary.

In the case of rotation ($\CKL > 0$), the parameters $\Omega_e$, $\CKL$, and
$\Theta$ all change on a timescale which is long compared to individual KL
cycles.  It is therefore possible to assume that $\Omega_e$, $\CKL$, and
$\Theta$ are all approximately constant over individual KL oscillations and
only examine the long-term changes to these parameters.  In this
approximation the system remains integrable with new integrals of motion.
Due to the integrability of the system the variations in $\CKL$, $\Theta$,
and $\Omega_e$ are all strictly periodic.  In this section we derive the
period of these EKM oscillations.  We note that there is a related
octupole-order dynamical phenomenon in which nearly coplanar orbits at high
eccentricity can undergo a flip by rolling over its major axis.  An analysis
of this phenomenon, including a derivation of the timescale, can be found in
\citet{li14a}.

\subsection{Equations of motion and integrals of motion}
\label{sec:octeom}

Since energy is conserved, the quadrupole order term of the Hamiltonian,
$\Hquad$, is also conserved in the time-averaged behavior of the system.
This implies that the relationship between $\CKL$, $\Theta$, and $\Hhatquad$
in equation~(\ref{eq:librationconst}) remains valid and that the quantity
\begin{equation}
\label{eq:phi}
\phi_q \equiv \CKL + \frac{1}{2} \Theta
\end{equation}
is a constant of motion. 

It is convenient to work with the parameter $\epsoct$, which measures the
relative size of the octupole order term of the Hamiltonian to the
quadrupole order term.  The parameter $\epsoct$ is conventionally defined as
\begin{equation}
\label{eq:epsoct_def}
\epsoct \equiv \frac{e_2}{1 - e_2^2} \frac{a_1}{a_2}.
\end{equation}
Some authors have added a mass term \citep[e.g.,][]{naoz13a} to capture the
fact that the octupole term is zero and EKM oscillations do not occur for an
equal mass inner binary.  However, because we are working exclusively in the
test particle limit we do not do so here. 

Following \citet{katz11}, the long-term evolution in $\Omega_e$
and $\Theta$ are given by
\begin{equation}
\label{eq:ekmeom_Omega}
\frac{d \Omega_e}{d\tau} = \Theta \left( \frac{6 E(x) - 3 K(x)}{4 K(x)}
\right),
\end{equation}
\begin{equation}
\label{eq:ekmeom_Theta}
\frac{d \Theta}{d\tau} = - \frac{15 \pi \epsoct}{64 \sqrt{10}}
\frac{\sqrt{\Theta} \sin \Omega_e}{K(x)} \left( 4 - 11 \CKL \right) \sqrt{6 +
4 \CKL},
\end{equation}
where $K(x)$ and $E(x)$ are complete elliptic functions of the first and
second kind, respectively, 
\begin{equation}
x(\CKL) \equiv \frac{3(1 - \CKL)}{3 + 2 \CKL},
\end{equation}
and the time coordinate has been scaled to the KL period during a flip:
\begin{equation}
\tau = \frac{t}{\tklflip}.
\end{equation}
\citet{katz11} also derive another integral of motion,
\begin{equation}
\label{eq:chi}
\chi \equiv F(\CKL) - \epsoct \cos \Omega_e,
\end{equation}
where the function $F(\CKL)$ is defined to be
\begin{equation}
F(\CKL) \equiv \frac{32 \sqrt{3}}{\pi} \int_{x(\CKL)}^1 \frac{K(\eta) - 2
E(\eta)}{(41 \eta - 21) \sqrt{2 \eta + 3}} \, d\eta.
\end{equation}

Although there are two integrals of motion, $\phi_q$ and $\chi$, they are
not sufficient to completely describe the dynamical behavior of the triple.
This is because $\epsoct$ carries dynamical information as well, most
importantly whether or not flips are possible.  The dynamical significance
of $\epsoct$ can be seen from the fact that $\epsoct$ enters into the
definition of $\chi$.  Thus, in the octupole case there are three
independent parameters describing the system as opposed to the case of
quadrupole-order KL oscillations in which there are only two.

\subsection{The period of EKM oscillations}
\label{subsec:ekmperiod}

In the case of EKM oscillations it is easier to derive their period directly
from the equations of motion rather than from action angle variables.  We
have from equation~(\ref{eq:phi}) that
\begin{equation}
\frac{d\CKL}{d\tau} = - \frac{1}{2} \frac{d \Theta}{d\tau},
\end{equation}
so the period may be written
\begin{equation}
\tauekm = \oint \, d\tau = \oint \frac{d \CKL}{\dot{C}_{\KL}}.
\end{equation}
Substituting equation~(\ref{eq:ekmeom_Theta}) we find
\begin{multline}
\label{eq:ekm_period_step1}
\tauekm = \oint \frac{128 \sqrt{10}}{15 \pi \epsoct}
\frac{K(x)}{\sqrt{2(\phi_q - \CKL)} \sin \Omega_e} 
\\
\times \frac{1}{(4 - 11 \CKL) \sqrt{6 + 4 \CKL}} \, d\CKL.
\end{multline}
To write $\Omega_e$ in terms of $\CKL$, we note that equation~(\ref{eq:chi})
implies that
\begin{equation}
\label{eq:sinOmega}
\sin \Omega_e = \sqrt{1 - \left( \frac{\chi - F(\CKL)}{\epsoct} \right)^2}.
\end{equation}
Substituting equation~(\ref{eq:sinOmega}) into
equation~(\ref{eq:ekm_period_step1}) and explicitly writing the limits of
the integral yields
\begin{multline}
\label{eq:tekm}
\tauekm = \frac{256 \sqrt{10}}{15 \pi \epsoct} \int_{\CKLmin}^{\CKLmax}
\frac{K(x)}{\sqrt{2(\phi_q - \CKL)} \left(4 - 11 \CKL \right)}
\\
\times \left[\left(1 - \frac{(\chi - F(\CKL))^2}{\epsoct^2} \right) \left(6 +
4 \CKL \right)\right]^{-\frac{1}{2}} \, d \CKL.
\end{multline}
The upper limit of the integral can be deduced by noting that $\CKL$ is
maximized when $\Theta$ is minimized and that $\Theta = 0$ during a flip.
We therefore have
\begin{equation}
\label{eq:CKLmax}
\CKLmax = \phi_q.
\end{equation}
The lower limit is more subtle.  It is clear from
equation~(\ref{eq:ekmeom_Theta}) that $\Theta$ is maximized when $\sin
\Omega_e = 0$.  This implies from equation~(\ref{eq:chi}) that 
\begin{equation}
\label{eq:CKLmin}
F(\CKLmin) = \chi \pm \epsoct.
\end{equation}
To decide whether to take the plus or minus sign, we must solve both for
$\CKL$ and then take the value of $\CKL$ which is less than $\CKLmax$.  This
equation can then be used to solve for $\CKLmin$ numerically.  Together,
equations~(\ref{eq:tekm}),~(\ref{eq:CKLmax}), and~(\ref{eq:CKLmin}) can be
used to solve for the period of EKM oscillations exactly.

\subsection{Parameter space of the EKM}
\label{subsec:ekmparam}

As in the quadrupole case we first explore over what region of parameter
space EKM oscillations with flips may occur.  We then determine the
variation in $\tekm$ over this range of parameter space.  Unfortunately, the
parameter space cannot be mapped quite as easily as in the case of
quadrupole KL oscillations because there are now three parameters describing
the system instead of two: $\phi_q$, $\chi$, and $\epsoct$.  As such, we
explore parameter space for two choices of $\epsoct$: $\epsoct = 10^{-3}$
and $\epsoct = 10^{-2}$.  Strong octupole-order effects occur in many triple
systems with $\epsoct = 10^{-2}$, but these effects are much weaker for most
triples when $\epsoct = 10^{-3}$ \citep[e.g.,][]{lithwick11}.

To determine the boundaries of the parameter space of spin flips we first
recall that $0 \leq \Theta \leq 1$, and for rotation $0 \leq \CKL \leq 1$
(which is the only case we are considering to octupole order).  The
occurrence of a spin flip is equivalent to having $\Theta = 0$, and hence
during a flip $\CKL = \phi_q$.  Since $\cos \Omega_e$ is bounded by $\pm1$,
we then have the following constraint:
\begin{equation}
F(\phi_q) - \epsoct \leq \chi \leq F(\phi_q) + \epsoct.
\end{equation}
The parameter space can be divided into two regions based on the maximum of
the function $F(\phi_q)$.  This maximum can be found by solving
$K(x_{\crit}) = 2 E(x_{\crit})$ for $x_{\crit}$, which yields $x_{\crit}
\approx 0.826$, and then calculating
\begin{equation}
\phiqcrit = \frac{3(1 - x_{\crit})}{3 + 2x_{\crit}} \approx 0.112.
\end{equation}
Now, $\phi_q$ cannot be arbitrarily large because $F(\phi_q)$ diverges at
$\phi_q = 4/11$.  Thus we have
\begin{equation}
\phi_q < \frac{4}{11}.
\end{equation}
Since, for $\phi_q < \phiqcrit$, $F(\phi_q)$ cannot be less than zero, this
then implies a constraint on $\chi$:
\begin{equation}
\chi \geq \epsoct \quad (\phi_q < \phiqcrit).
\end{equation}
Finally, the above relation implies that 
\begin{equation}
F(\phi_{q, \min}) = \epsoct.
\end{equation}

\begin{figure*}
\centering
\includegraphics[width=16cm]{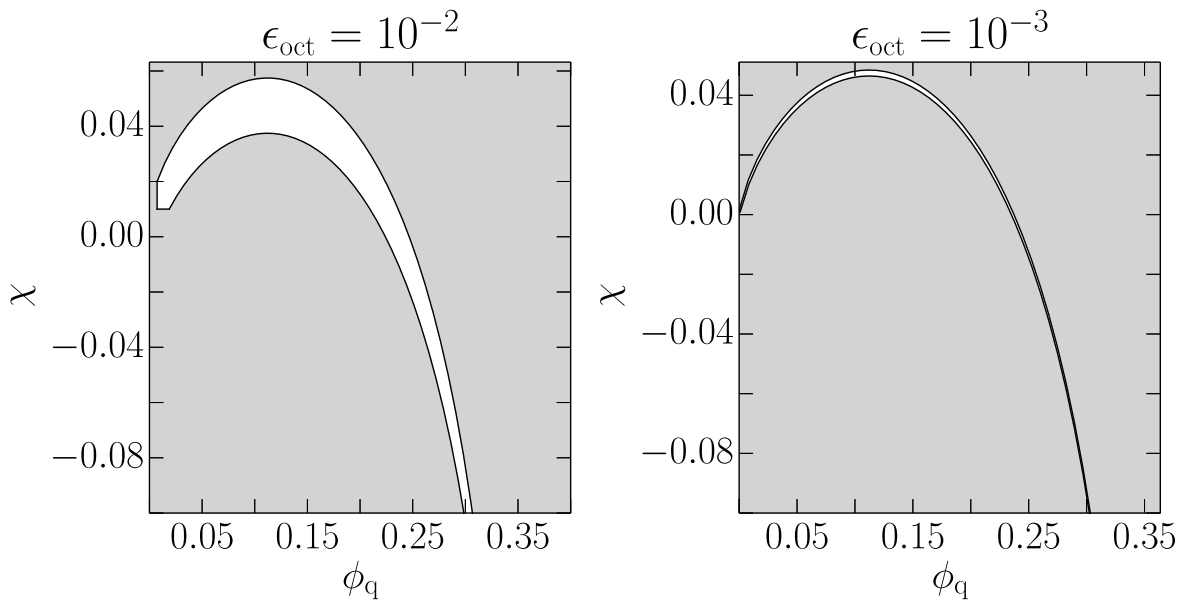}

\caption{Parameter space where EKM oscillations with flips are possible for
two choices of $\epsoct$.  We only explore the parameter space where
individual KL cycles are rotating instead of librating (i.e., $\CKL > 0$),
as librating cycles cannot be correctly analyzed using this technique of
averaging over individual KL oscillations.  At smaller values of $\epsoct$
the area of parameter space where rotating flips are possible shrinks about
the line $\chi = F(\phi)$.}

\label{fig:ekmpspace}
\end{figure*}

\begin{figure*}
\centering
\includegraphics[width=16cm]{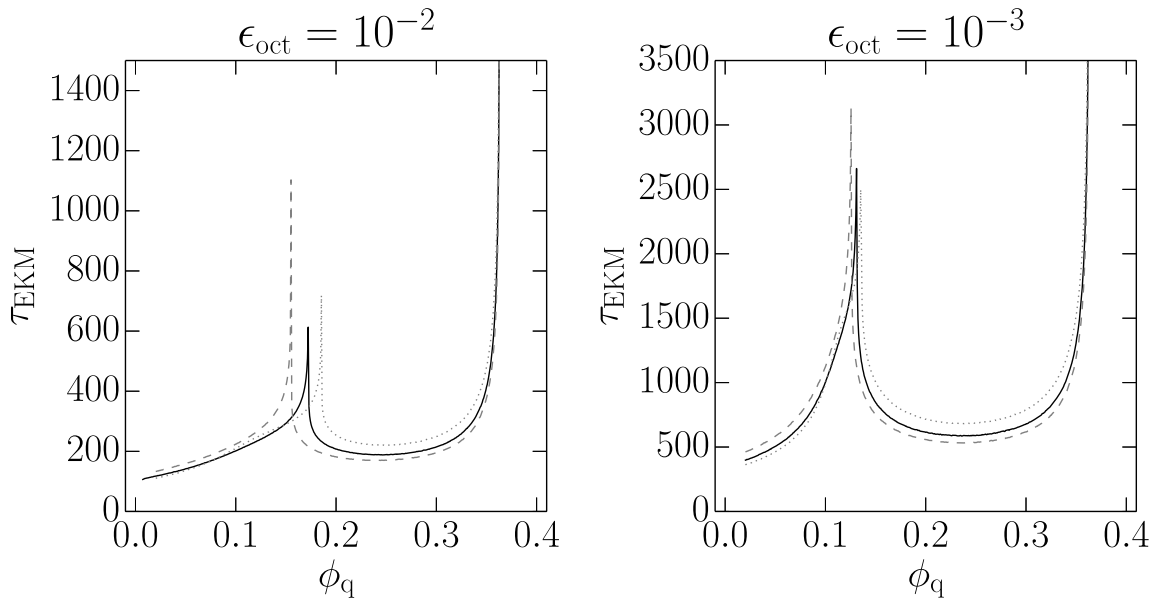}

\caption{The period of EKM oscillations with flips as a function of $\phi_q$
for three choices of $\chi$.  The solid line is given by the choice $\chi =
F(\phi_q)$, the dashed line by $\chi = F(\phi_q) - \epsoct / 2$, and the
dotted line by $\chi = F(\phi_q) + \epsoct / 2$.  Except very near the two
singularities, the period of EKM oscillations does not vary by more than a
factor of a few.  Over a broad range of parameter space EKM oscillations
have similar timescales.}

\label{fig:phidep}
\end{figure*}

Taken together, these relations bound the parameter space over which flips
are possible.  The resulting maps of parameter space for $\epsoct = 10^{-3}$
and $\epsoct = 10^{-2}$ are shown in Figure~\ref{fig:ekmpspace}.  Because
the parameter space over which flips occur is somewhat narrow and the
dependence of $\tau_{\ekm}$ on $\phi_q$ is fairly complicated we do not show
contours as we did at quadrupole order in Fig.~\ref{fig:pspace}.  Instead,
we show $\tau_{\ekm}$ as a function of $\phi_q$ with the choice of $\chi =
F(\phi_q)$ and $\chi = F(\phi_q) \pm \epsoct / 2$ in Fig.~\ref{fig:phidep}.
The timescale for EKM oscillations depends most sensitively on $\phi_q$.
The timescale has two singularities: one at the maximum value of $\phi_q$ of
4/11, and another which is dependent on the choice of $\chi$, but is near
$\phiqcrit$.  Except very close to these singularities, the period of EKM
oscillations does not vary by more than a factor of a few.  Thus, over a
broad range of parameter space EKM oscillations have similar timescales.
The existence of these singularities, however, does imply that $\tekm$ has
some dependence on, e.g., the initial inclination as was found by
\citet{teyssandier13}.

\subsection{The dependence on $\epsoct$}
\label{subsec:epsoct_dependence}

If the constants $\phi_q$ and $\chi$ are held fixed and $\epsoct$ is varied,
how does the period of EKM oscillations vary?  Equation~(\ref{eq:tekm})
exhibits a $\epsoct^{-1}$ dependence in the coefficient before the integral,
so it is tempting to conclude that the timescale for EKM oscillations scales
as $\epsoct^{-1}$.  This conclusion has been asserted in several studies in
the literature, but we show here that it is incorrect.  The integral in
equation~(\ref{eq:tekm}) in fact exhibits a $\epsoct^{-1/2}$ dependence.

To determine this dependence we first note that for EKM oscillations to
occur, in general $\CKL \ll 1$.  This then implies that $x$ is very close to
unity, so we may write $x = 1 - \varepsilon$, where $\varepsilon \ll 1$.
For values of $x$ very close to unity, the complete elliptic integral of the
first kind may be approximated
\begin{equation}
\label{eq:Kapprox}
K(1 - \varepsilon) \simeq - \frac{1}{2} \ln \varepsilon,
\end{equation}
and the complete elliptic integral of the second kind is approximated by
$E(1 - \varepsilon) \simeq 1$.  We note that the coefficient in
equation~(\ref{eq:Kapprox}) is off by several tens of percent for realistic
values of $\varepsilon$, but the important feature of this approximation is
that it carries the correct dependence on $\varepsilon$.  The function
$F(\CKL)$ can then be approximated as
\begin{equation}
F(\CKL) \simeq -\frac{8}{5 \pi} \sqrt{\frac{3}{5}} \int_0^{\varepsilon}
\left(\frac{1}{2} \ln \left( \varepsilon^{\prime} \right) + 2 \right) \,
d\varepsilon^{\prime}.
\end{equation}
Now, because we are integrating over a small range, the integral can then be
approximated as
\begin{equation}
F(\CKL) \simeq -\frac{8}{5 \pi} \sqrt{\frac{3}{5}} \left(\frac{1}{2} \ln
\left( \frac{\varepsilon^{\prime}}{2} \right) + 2 \right) \varepsilon
\end{equation}
Furthermore, we note that
\begin{equation}
\varepsilon \simeq \frac{2}{3} \CKL
\end{equation}
so we finally have
\begin{equation}
\label{eq:Fapprox}
F(\CKL) \simeq -\frac{16}{15 \pi} \sqrt{\frac{3}{5}} \CKL \left( \frac{1}{2}
\ln \left( \frac{\CKL}{3} \right) + 2 \right).
\end{equation}

Let us now consider the lower limit of the integral in
equation~(\ref{eq:tekm}).  For simplicity, let us for the time being
restrict ourselves to the locus $\chi = F(\phi_q)$ since here flips occur
for arbitrarily small values of $\epsoct$.  We then have
\begin{equation}
F(\CKLmin) = F(\phi_q) - \epsoct.
\end{equation}
Now, the approximation in equation~(\ref{eq:Fapprox}) may be written more
simply as $F(\CKL) \sim k \CKL$, where $k$ is a parameter that has only a
sub-linear dependence on $\CKL$.  For small $\CKL$, then, the function $F$
is nearly linear in $\CKL$.  This then implies that for points on the locus
we are considering
\begin{equation}
\label{eq:deltaCKLapprox}
\phi_q - \CKLmin \sim \frac{\epsoct}{k}.
\end{equation}
This then means that the width over which we are integrating, $\Delta \CKL$
is proportional to $\epsoct$ since
\begin{equation}
\Delta \CKL \equiv \CKLmax - \CKLmin = \phi_q - \CKLmin \sim \epsoct.
\end{equation}

Let us now consider the various terms of the integrand of
equation~(\ref{eq:tekm}).  We have already seen that because $x$ is close to
unity, $K(x) \sim \ln(\CKL/3)$.  This term is sublinear so we ignore it.
The $(4 - 11 \CKL)$ term reduces to 4, and similarly the $\sqrt{6 + 4 \CKL}$
term reduces to $\sqrt{6}$.  The term $\sqrt{2(\phi_q - \CKL)}$ reduces by
equation~(\ref{eq:deltaCKLapprox}) to $\sim$$\sqrt{2 \epsoct}$.  This leaves
only the $\sin \Omega_e$ term.  Now, if $\phi_q - \CKL \sim \epsoct$ and $F$
is approximately linear in this limit, it must be the case that
\begin{equation}
F(\phi_q) - \CKL = \chi - \CKL \sim \epsoct.
\end{equation}
Comparing this to equation~(\ref{eq:sinOmega}), we find that to lowest
order, $\sin \Omega_e$ does not exhibit any dependence on $\epsoct$.  It is
straightforward to verify this claim numerically.

Putting these results together, we find that the only dependencies on
$\epsoct$ in the integral come from the width of integration (which yields a
dependence of $\epsoct$), and from the term $1/\sqrt{2(\phi_q - \CKL)}$
(which yields a dependence of $\epsoct^{-1/2}$).  Since the integral has a
coefficient of $\epsoct^{-1}$, this then implies that the overall dependence
of the period of the EKM is
\begin{equation}
\tauekm \sim \frac{1}{\sqrt{\epsoct}}.
\end{equation}

\begin{figure}
\centering
\includegraphics[width=8cm]{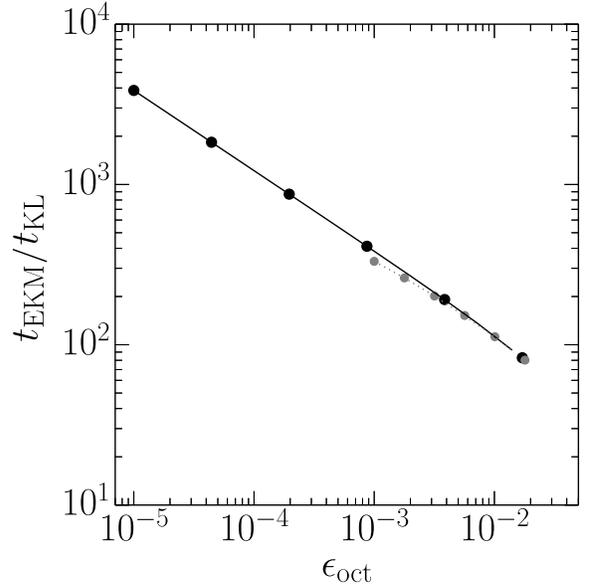}

\caption{The period of the EKM relative to the period of KL oscillations as
a function of $\epsoct$ calculated analytically using
equation~(\ref{eq:tekm}) (lines) and by integrating the secular equations of
motion (points).  The timescale for the EKM is almost exactly proportional
to $\epsoct^{-1/2}$ and there is excellent agreement between the secular and
analytic calculations.  We show this relationship for an arbitrary choice of
$\phi_q = 0.015$ and two choices of $\chi = F(\phi_q)$ (black line and
points), and $\chi = F(\phi_q) + 9 \times 10^{-4}$ (gray dotted line and
points).  For $\chi = F(\phi_q)$ flips are possible at arbitrarily small
values of $\epsoct$, whereas for $\chi = F(\phi_q) + 9 \times 10^{-4}$ flips
are only possible for values of $\epsoct > 9 \times 10^{-4}$.  The
relationship between $\tekm$ and $\epsoct$ becomes slightly shallower near
this critical value of $\epsoct$.  Note that $\epsoct$ cannot exceed $\chi$.
Although flips occur at larger values of $\epsoct$, the evolution is no
longer integrable because the inner binary switches between rotation and
libration.  In this regime the timescale for flips steepens as a function of
$\epsoct$, although there is no longer a simple relationship between the two
because the evolution becomes essentially chaotic.}

\label{fig:epsdep}
\end{figure}

We demonstrate this dependence explicitly in Fig.~\ref{fig:epsdep} by
numerically calculating the period using equation~(\ref{eq:tekm}) for fixed
values of $\phi_q$ and $\chi$ but over a range of $\epsoct$.  We have
compared these values with the periods obtained by integrating the secular
equations of motion directly and find excellent agreement. 

By combining this result with the numerical coefficient of
equation~(\ref{eq:tekm}) we find that
\begin{equation}
\label{eq:ekm_timescale}
t_{\ekm} \sim \frac{256 \sqrt{10}}{15 \pi \sqrt{\epsoct}} \tklflip.
\end{equation}
During a single EKM cycle the inner binary will undergo two flips, so the
flip timescale is half this value.  The flip timescale can then be obtained
by substituting for $\tklflip$, taking $\Theta = 0$ in
equation~(\ref{eq:approx}), and noting from equation~(\ref{eq:phi}) that
$\CKL = \phi_q$ when $\Theta = 0$, 
\begin{equation}
\label{eq:flip_timescale}
t_{\textrm{flip}} \sim \frac{256}{9 \pi^2 \sqrt{15 \epsoct}} \left(1 +
\frac{m_1}{m_3}\right) \left(\frac{P_{\out}^2}{P_{\in}}\right) \left(1 -
e_2^2\right)^{3/2} \ln \left(\frac{9.42}{\phi_q}\right).
\end{equation}
Over most of parameter space this expression is valid to within a factor of
a few.  For extremely large values of $\epsoct$ ($\epsoct \sim 0.1$) our
numerical experiments demonstrate that the dependence of $\tekm$ on
$\epsoct$ steepens and this expression overpredicts the timescale for flips,
but in this limit non-secular effects become important, so the above
analysis does not apply \citep[e.g.,][]{antonini12, katz12, seto13,
antonini14, bode14, antognini14}.  Moreover, the above analysis also
requires individual KL oscillations to be short relative to the EKM cycle.
If this is not the case, then resonances between the quadrupole and octupole
order terms can induce chaotic variation of the orbital eccentricity
\citep{li14b}.  Thus equation~(\ref{eq:ekm_timescale}) should be taken as a
rough upper limit on $\tekm$. 

\section{Conclusions}
\label{sec:conclusions}

Using action angle variables we have derived the period of KL oscillations
at quadrupole order and in the test particle limit
(equation~\ref{eq:kl_period}).  From the exact period we have derived the
timescale for KL oscillations.  We have explored the full range of parameter
space over which KL oscillations are possible and found that except very
near the boundary between rotation and libration ($|\CKL| \ll 1$) the period
of KL oscillations does not vary by more than a factor of a few from the
derived timescale (Fig.~\ref{fig:pspace}).  By employing several
approximations in the high-inclination, low initial eccentricity limit we
have found a function that matches the true KL period to within 2\% for
triples for which $e_1 \lesssim 0.3$ and $i \gtrsim 60^{\circ}$
(equation~\ref{eq:approx}).

The strength of KL oscillations varies due to the octupole term of the
Hamiltonian.  We average over individual KL cycles to calculate the period
of EKM oscillations, and hence, the timescale for spin flips to occur.  We
map the parameter space over which spin flips occur
(Fig.~\ref{fig:ekmpspace}) and show that apart from near two singularities
where spin flips do not occur, the timescale for EKM oscillations does not
vary by more than a factor of a few (Fig.~\ref{fig:phidep}).  Finally, we
show numerically and analytically that the dependence of $\epsoct$ on the
timescale for EKM oscillations is $\epsoct^{-1/2}$ (Fig.~\ref{fig:epsdep})
in contrast to previous studies.  We provide the EKM timescale in
equation~(\ref{eq:ekm_timescale}) and the timescale for flips in
equation~(\ref{eq:flip_timescale}).

\section*{Acknowledgements}

The author thanks Todd Thompson for many helpful suggestions and a close
reading of the manuscript.  The author thanks Christopher Kochanek, Ondrej
Pejcha, Boaz Katz, Yoram Lithwick, Renu Malhotra, Cristobal Petrovich, and
Benjamin Shappee for their comments.  The author also thanks Scott Tremaine
for pointing out that equation~(\ref{eq:loopintegral}) is an elliptic
integral.  This paper made use of \textsc{MatPlotLib} \citep{hunter07} and
\textsc{IPython} \citep{perez07}.  This research was supported by the
National Science Foundation under NSF AST Award No.~1313252.

\bibliographystyle{mn2e}
\bibliography{refs}

\begin{thebibliography}{49}
\expandafter\ifx\csname natexlab\endcsname\relax\def\natexlab#1{#1}\fi

\bibitem[{{Antognini} {et~al}\mbox{.}(2014){Antognini}, {Shappee}, {Thompson},
  \& {Amaro-Seoane}}]{antognini14}
{Antognini} J.~M., {Shappee} B.~J., {Thompson} T.~A., {Amaro-Seoane} P., 2014,
  \mnras, 439, 1079

\bibitem[{{Antonini} {et~al}\mbox{.}(2014){Antonini}, {Murray}, \&
  {Mikkola}}]{antonini14}
{Antonini} F., {Murray} N., {Mikkola} S., 2014, \apj, 781, 45

\bibitem[{{Antonini} \& {Perets}(2012)}]{antonini12}
{Antonini} F., {Perets} H.~B., 2012, \apj, 757, 27

\bibitem[{{Blaes} {et~al}\mbox{.}(2002){Blaes}, {Lee}, \& {Socrates}}]{blaes02}
{Blaes} O., {Lee} M.~H., {Socrates} A., 2002, \apj, 578, 775

\bibitem[{{Bode} \& {Wegg}(2014)}]{bode14}
{Bode} J.~N., {Wegg} C., 2014, \mnras, 438, 573

\bibitem[{Chenciner \& Montgomery(2000)}]{chenciner00}
Chenciner A., Montgomery R., 2000, Annals of Mathematics-Second Series, 152,
  881

\bibitem[{{Duquennoy} \& {Mayor}(1991)}]{duquennoy91}
{Duquennoy} A., {Mayor} M., 1991, \aap, 248, 485

\bibitem[{{Fabrycky} \& {Tremaine}(2007)}]{fabrycky07}
{Fabrycky} D., {Tremaine} S., 2007, \apj, 669, 1298

\bibitem[{{Ford} {et~al}\mbox{.}(2000){Ford}, {Kozinsky}, \& {Rasio}}]{ford00}
{Ford} E.~B., {Kozinsky} B., {Rasio} F.~A., 2000, \apj, 535, 385

\bibitem[{{Gould}(2011)}]{gould11}
{Gould} A., 2011, \apjl, 729, L23

\bibitem[{{Hamers} {et~al}\mbox{.}(2015){Hamers}, {Perets}, {Antonini}, \&
  {Portegies Zwart}}]{hamers15}
{Hamers} A.~S., {Perets} H.~B., {Antonini} F., {Portegies Zwart} S.~F., 2015,
  \mnras, 449, 4221

\bibitem[{{Harrington}(1968)}]{harrington68}
{Harrington} R.~S., 1968, \aj, 73, 190

\bibitem[{{Holman} {et~al}\mbox{.}(1997){Holman}, {Touma}, \&
  {Tremaine}}]{holman97}
{Holman} M., {Touma} J., {Tremaine} S., 1997, \nat, 386, 254

\bibitem[{Hunter(2007)}]{hunter07}
Hunter J.~D., 2007, Computing In Science \& Engineering, 9, 90

\bibitem[{{Katz} \& {Dong}(2012)}]{katz12}
{Katz} B., {Dong} S., 2012, ArXiv e-prints: 1211.4584

\bibitem[{{Katz} {et~al}\mbox{.}(2011){Katz}, {Dong}, \& {Malhotra}}]{katz11}
{Katz} B., {Dong} S., {Malhotra} R., 2011, Physical Review Letters, 107, 181101

\bibitem[{{Kozai}(1962)}]{kozai62}
{Kozai} Y., 1962, \aj, 67, 591

\bibitem[{{Li} {et~al}\mbox{.}(2014{\natexlab{a}}){Li}, {Naoz}, {Holman}, \&
  {Loeb}}]{li14b}
{Li} G., {Naoz} S., {Holman} M., {Loeb} A., 2014{\natexlab{a}}, \apj, 791, 86

\bibitem[{{Li} {et~al}\mbox{.}(2014{\natexlab{b}}){Li}, {Naoz}, {Kocsis}, \&
  {Loeb}}]{li14a}
{Li} G., {Naoz} S., {Kocsis} B., {Loeb} A., 2014{\natexlab{b}}, \apj, 785, 116

\bibitem[{{Li} {et~al}\mbox{.}(2015){Li}, {Naoz}, {Kocsis}, \& {Loeb}}]{li15b}
{Li} G., {Naoz} S., {Kocsis} B., {Loeb} A., 2015, ArXiv e-prints

\bibitem[{{Lidov}(1962)}]{lidov62}
{Lidov} M.~L., 1962, \planss, 9, 719

\bibitem[{{Lidov} \& {Ziglin}(1976)}]{lidov76}
{Lidov} M.~L., {Ziglin} S.~L., 1976, Celestial Mechanics, 13, 471

\bibitem[{{Lithwick} \& {Naoz}(2011)}]{lithwick11}
{Lithwick} Y., {Naoz} S., 2011, \apj, 742, 94

\bibitem[{{Mardling} \& {Aarseth}(1999)}]{mardling99}
{Mardling} R., {Aarseth} S., 1999, in NATO ASIC Proc. 522: The Dynamics of
  Small Bodies in the Solar System, A Major Key to Solar System Studies,
  {Steves} B.~A., {Roy} A.~E., eds., p. 385

\bibitem[{{Mardling} \& {Aarseth}(2001)}]{mardling01}
{Mardling} R.~A., {Aarseth} S.~J., 2001, \mnras, 321, 398

\bibitem[{{Michaely} \& {Perets}(2014)}]{michaely14}
{Michaely} E., {Perets} H.~B., 2014, \apj, 794, 122

\bibitem[{{Miller} \& {Hamilton}(2002{\natexlab{a}})}]{miller02a}
{Miller} M.~C., {Hamilton} D.~P., 2002{\natexlab{a}}, \apj, 576, 894

\bibitem[{{Miller} \& {Hamilton}(2002{\natexlab{b}})}]{miller02b}
{Miller} M.~C., {Hamilton} D.~P., 2002{\natexlab{b}}, \mnras, 330, 232

\bibitem[{{Naoz} \& {Fabrycky}(2014)}]{naoz14a}
{Naoz} S., {Fabrycky} D.~C., 2014, \apj, 793, 137

\bibitem[{{Naoz} {et~al}\mbox{.}(2011){Naoz}, {Farr}, {Lithwick}, {Rasio}, \&
  {Teyssandier}}]{naoz11}
{Naoz} S., {Farr} W.~M., {Lithwick} Y., {Rasio} F.~A., {Teyssandier} J., 2011,
  \nat, 473, 187

\bibitem[{{Naoz} {et~al}\mbox{.}(2013{\natexlab{a}}){Naoz}, {Farr}, {Lithwick},
  {Rasio}, \& {Teyssandier}}]{naoz13a}
{Naoz} S., {Farr} W.~M., {Lithwick} Y., {Rasio} F.~A., {Teyssandier} J.,
  2013{\natexlab{a}}, \mnras, 431, 2155

\bibitem[{{Naoz} {et~al}\mbox{.}(2012){Naoz}, {Farr}, \& {Rasio}}]{naoz12}
{Naoz} S., {Farr} W.~M., {Rasio} F.~A., 2012, \apjl, 754, L36

\bibitem[{{Naoz} {et~al}\mbox{.}(2013{\natexlab{b}}){Naoz}, {Kocsis}, {Loeb},
  \& {Yunes}}]{naoz13b}
{Naoz} S., {Kocsis} B., {Loeb} A., {Yunes} N., 2013{\natexlab{b}}, \apj, 773,
  187

\bibitem[{{Naoz} \& {Silk}(2014)}]{naoz14b}
{Naoz} S., {Silk} J., 2014, \apj, 795, 102

\bibitem[{{Pejcha} {et~al}\mbox{.}(2013){Pejcha}, {Antognini}, {Shappee}, \&
  {Thompson}}]{pejcha13}
{Pejcha} O., {Antognini} J.~M., {Shappee} B.~J., {Thompson} T.~A., 2013,
  \mnras, 435, 943

\bibitem[{{Perets} \& {Fabrycky}(2009)}]{perets09}
{Perets} H.~B., {Fabrycky} D.~C., 2009, \apj, 697, 1048

\bibitem[{P\'erez \& Granger(2007)}]{perez07}
P\'erez F., Granger B.~E., 2007, Computing in Science and Engineering, 9, 21

\bibitem[{{Petrovich}(2015)}]{petrovich15a}
{Petrovich} C., 2015, \apj, 799, 27

\bibitem[{{Raghavan} {et~al}\mbox{.}(2010){Raghavan}, {McAlister}, {Henry},
  {Latham}, {Marcy}, {Mason}, {Gies}, {White}, \& {ten
  Brummelaar}}]{raghavan10}
{Raghavan} D. {et~al.}, 2010, \apjs, 190, 1

\bibitem[{{Riddle} {et~al}\mbox{.}(2015){Riddle}, {Tokovinin}, {Mason},
  {Hartkopf}, {Roberts}, {Baranec}, {Law}, {Bui}, {Burse}, {Das}, {Dekany},
  {Kulkarni}, {Punnadi}, {Ramaprakash}, \& {Tendulkar}}]{riddle15}
{Riddle} R.~L. {et~al.}, 2015, \apj, 799, 4

\bibitem[{{Seto}(2013)}]{seto13}
{Seto} N., 2013, Physical Review Letters, 111, 061106

\bibitem[{{Shappee} \& {Thompson}(2013)}]{shappee13b}
{Shappee} B.~J., {Thompson} T.~A., 2013, \apj, 766, 64

\bibitem[{{Teyssandier} {et~al}\mbox{.}(2013){Teyssandier}, {Naoz},
  {Lizarraga}, \& {Rasio}}]{teyssandier13}
{Teyssandier} J., {Naoz} S., {Lizarraga} I., {Rasio} F.~A., 2013, \apj, 779,
  166

\bibitem[{{Thompson}(2011)}]{thompson11}
{Thompson} T.~A., 2011, \apj, 741, 82

\bibitem[{{Tokovinin}(2014)}]{tokovinin14}
{Tokovinin} A., 2014, \aj, 147, 87

\bibitem[{{{\v S}uvakov} \& {Dmitra{\v s}inovi{\'c}}(2013)}]{suvakov13}
{{\v S}uvakov} M., {Dmitra{\v s}inovi{\'c}} V., 2013, Physical Review Letters,
  110, 114301

\bibitem[{{Wen}(2003)}]{wen03}
{Wen} L., 2003, \apj, 598, 419

\bibitem[{{Wu} \& {Murray}(2003)}]{wu03}
{Wu} Y., {Murray} N., 2003, \apj, 589, 605

\bibitem[{{Wu} {et~al}\mbox{.}(2007){Wu}, {Murray}, \& {Ramsahai}}]{wu07}
{Wu} Y., {Murray} N.~W., {Ramsahai} J.~M., 2007, \apj, 670, 820

\end{thebibliography}

\end{document}